\documentclass[pra,twocolumn,graphicx]{revtex4}
\usepackage{amssymb}
\usepackage{epsfig,amsmath}

\begin{document}

\title{High-harmonic generation in diatomic molecules: a quantum-orbit
analysis of the interference patterns}
\author{C. Figueira de Morisson Faria}
\affiliation{Department of Physics and Astronomy, University College London, Gower
Street, London WC1E 6BT, United Kingdom}
\date{\today}

\begin{abstract}
We perform a detailed analysis of high-order harmonic generation in
diatomic molecules within the strong-field approximation, with
emphasis on quantum-interference effects. Specifically, we
investigate how the different types of electron orbits, involving
one or two centers, affect the interference patterns in the spectra.
We also briefly address the influence of the choice of gauge, and of
the initial and final electronic bound states on such patterns. For
the length-gauge SFA and undressed bound states, there exist
additional terms, which can be interpreted as potential energy
shifts. If, on the one hand, such shifts alter the potential
barriers through which the electron initially tunnels, and may lead
to a questionable physical interpretation of the features
encountered, on the other hand they seem to be necessary in order to
reproduce the overall maxima and minima in the spectra. Indeed, for
dressed electronic bound states in the length gauge, or undressed
bound states in the velocity gauge, for which such shifts are
absent, there is a breakdown of the interference patterns. In order
to avoid such a problem, we provide an alternative pathway for the
electron to reach the continuum, by means of an additional
attosecond-pulse train. A comparison of the purely monochromatic
case with the situation for which the attosecond pulses are present
suggests that the patterns are due to the interference between the
electron orbits which finish at different centers, regardless of
whether one or two centers are involved.
\end{abstract}

\maketitle

\section{Introduction}

In the past few years, high-order harmonic generation (HHG) and
above-threshold ionization (ATI) from aligned molecules in strong laser
fields of femtosecond duration have proven to be a powerful tool for
resolving, or even controlling, processes in the subfemtosecond and
subangstrom scale. For instance, one may employ HHG and ATI in the
tomographic reconstruction of molecular orbitals \cite{Itatani}, and in the
attosecond probing of dynamic changes in molecules \cite{attomol}.

This is possible due to the fact that the physical mechanisms governing both
phenomena take place in a fraction of the laser period, i.e., within
hundreds of attoseconds \cite{Scrinzi2006}, and involve the recombination or
the elastic scattering of an electron with its parent molecule \cite{tstep}.
Thereby, high-order harmonics or high-energy photoelectrons, respectively,
are generated. Thus, the spectral features are highly dependent on the
spatial configuration of the ions with which the electron rescatters or
recombines, and yield patterns which are characteristic of the molecule.
Furthermore, they also depend on the alignment angle of the molecule, with
respect to the laser-field polarization.

Due to their simplicity, in particular diatomic molecules have been
investigated, and minima and maxima have been encountered in their HHG and
ATI spectra. Such patterns have been observed both theoretically \cite%
{TDSE2005,doubleslit,KB2005,KBK98,MBBF00,Madsen,Usach2006,Usachenko,HenrikDipl,Kansas,PRACL2006,moreCL,DM2006,HBF2007}
and experimentally \cite{interfexp}, and have been attributed to the
interference between high- order harmonics or photoelectrons generated at
different centers in the molecule. They are, in a sense, the microscopic
counterpart of those obtained in a double-slit experiment. Furthermore, the
energy positions of the maxima and minima depend on the alignment angle and
on the internuclear distance and, additionally, reflect the bonding or
antibonding nature of the highest occupied molecular orbitals in question
\cite{TDSE2005,doubleslit,KB2005}. Such features have been studied either
numerically, by solving the time-dependent Schr\"{o}dinger equation (TDSE)
\cite{TDSE2005,doubleslit,KB2005}, or semi-analytically, by employing the
strong-field approximation (SFA) \cite%
{KBK98,MBBF00,Madsen,Usach2006,Usachenko,HenrikDipl,Kansas,PRACL2006,moreCL,DM2006,HBF2007}%
. In particular, in \cite{KB2005}, the contributions to the yield from each
molecular center have been singled out within a TDSE computation. Therein,
it has been explicitly shown that the maxima and minima in the spectra are
obtained due to the interference between contributions from different
centers, in agreement with the double-slit model in \cite{doubleslit,KB2005}.

Specifically in the SFA framework, the transition amplitude can be written
as a multiple integral, with a semiclassical action and slowly-varying
prefactors. The structure of the molecule (and thus its double-slit
character) can be either incorporated in the prefactors or in the action.
The latter approach also takes into account processes in which an electron
rescatters or recombines with a center in the molecule different from the
site of its release \cite{KBK98,HenrikDipl,Usach2006,PRACL2006,HBF2007}.

An open question is, however, which types of electron orbits are responsible
for specific interference features. For instance, are the dips and maxima
originated by the interference between orbits in which the electron leaves
and returns to the same center (regardless of which), or between those in
which the electron leaves one atom and recombines with the other? On the
other hand, it could also be that the interference patterns result from the
combined effect of all such orbits, and one is not able to attribute them to
specific sets.

Since the strong-field approximation is a semi-analytical method, and allows
an immediate association with the classical orbits of an electron returning
to its parent molecule, it appears to be an ideal tool for tackling this
problem. This approximation, however, possesses several drawbacks. First,
the SFA is gauge dependent, which leads to different pre-factors and action,
depending on whether the velocity gauge or the length gauge is taken.
Second, even for a specific gauge, the precise expressions for the
pre-factors are not agreed upon, and some of them lead to interference
patterns which are in disagreement with the experiments, and with results
from other methods \cite{KB2005,JMOCL2006}.

The main objective of this work is to make a detailed assessment of the
contributions and relevance of the different types of recombination
scenarios to the above-mentioned interference patterns, within the
Strong-Field Approximation, for a diatomic molecule. Throughout the paper,
we approximate the strong, infra-red field by \ a linearly polarized
monochromatic wave $\mathbf{E}(t)=E_{0}\sin (\omega t)\mathbf{e}_{x}$, and
consider a linear combination of atomic orbitals (LCAO approximation). In
general, we consider that the electron reaches the continuum by tunneling
ionization. As an exception, however, we also take an attosecond-pulse train
superposed to the monochromatic wave \cite%
{atthhg2004,attati2005,FSVL2006,FS2006}. The attosecond-pulse train provides
an additional pathway for the electron to reach the continuum, and,
recently, has proven to be a powerful tool in order to control high-order
harmonic generation \cite{atthhg2004,FSVL2006,FS2006} and above-threshold
ionization \cite{attati2005,FSVL2006}. In the context of the present work,
it is a convenient way to avoid problems related to spurious
potential-energy shifts. These shifts are present in the length gauge and
artificially modify the potential barrier through which the electron
tunnels. Hence, they may lead to a questionable physical interpretation,
what the relevance of the different sets of orbits to the patterns concerns.

This paper is organized as follows. In Sec. \ref{transampl}, we discuss the
strong-field approximation transition amplitude for high-order harmonic
generation, starting from the general expressions (Sec. \ref{general}), and,
subsequently, addressing the specific situation of a diatomic molecule in
the LCAO approximation (Sec. \ref{diatomic}). Thereby, we consider the
situation for which the structure of the molecule is either incorporated in
the pre-factor, or in the semiclassical action, in the presence and absence
of the attosecond-pulse train. When discussing the former case, we emphasize
the role of the overlap integrals, in which the dipole moment and the atomic
wave functions are localized at different centers in the molecule. In the
latter case, we follow the model in Ref. \cite{PRACL2006}, for a purely
monochromatic driving field, and our previous work \cite{FSVL2006} when the
attosecond pulses are present, very closely. Subsequently (Sec. \ref{results}%
), we investigate the interference patterns. Finally, in Sec. \ref{concl} we
summarize the paper and state our main conclusions.

\section{Transition amplitudes}

\label{transampl}

\subsection{General expressions}

\label{general}

In general, the strong-field approximation (SFA) consists in neglecting the
influence of the laser field when the electron is bound, the atomic or
molecular binding potential when the electron is in the continuum and the
internal structure of the system in question, i.e., contributions from its
excited bound states. The SFA transition amplitude for high-order harmonic
generation reads, in the specific formulation of Ref. \cite{hhgsfa} and in
atomic units,
\begin{eqnarray}
b_{\Omega } &\hspace{-0.1cm}=\hspace*{-0.1cm}&i\int_{-\infty }^{\infty }%
\hspace*{-0.5cm}dt\int_{-\infty }^{t}~\hspace*{-0.5cm}dt^{\prime }\int
d^{3}pd_{\mathrm{rec}}^{\ast }(\mathbf{\tilde{p}}(t))d_{\mathrm{ion}}(%
\mathbf{\tilde{p}}(t^{\prime }))  \notag \\
&&\exp [iS(t,t^{\prime },\Omega ,\mathbf{k})]+c.c.,  \label{amplhhg}
\end{eqnarray}%
with the action
\begin{equation}
S(t,t^{\prime },\Omega ,\mathbf{p})=-\frac{1}{2}\int_{t^{\prime }}^{t}[%
\mathbf{p}+\mathbf{A}(\tau )]^{2}d\tau -I_{p}(t-t^{\prime })+\Omega t
\label{actionhhg}
\end{equation}%
and the prefactors $d_{\mathrm{rec}}(\mathbf{\tilde{p}}(t))=\left\langle
\mathbf{\tilde{p}}(t)\right\vert \mathcal{O}_{\mathrm{dip}}.\mathbf{e}%
_{x}\left\vert \psi _{0}\right\rangle $ and $d_{\mathrm{ion}}(\mathbf{\tilde{%
p}}(t^{\prime }))=\left\langle \mathbf{\tilde{p}}(t^{\prime })\right\vert H_{%
\mathrm{int}}\mathbf{(}t^{\prime }\mathbf{)}\left\vert \psi
_{0}\right\rangle .$ In the above equations, $\mathcal{O}_{\mathrm{dip}}$, $%
\mathbf{e}_{x}$, $H_{\mathrm{int}}\mathbf{(}t^{\prime }\mathbf{),}$ $I_{p},$
and $\Omega $ denote the dipole operator, the laser-polarization vector, the
interaction with the field, the ionization potential, and the harmonic
frequency, respectively. The explicit expressions for $\mathbf{\tilde{p}}(t)$
are gauge dependent, and will be given below. The above-stated equation
describes a physical process in which an electron, initially in a field-free
bound-state $\left\vert \psi _{0}\right\rangle $, is coupled to a Volkov
state $\left\vert \mathbf{\tilde{p}}(t^{\prime })\right\rangle $ by the
interaction $H_{\mathrm{int}}\mathbf{(}t^{\prime }\mathbf{)}$ of the system
with the field. Subsequently, it propagates in the continuum and is driven
back towards its parent ion, with which it recombines at a time $t$,
emitting high-harmonic radiation of frequency $\Omega .$ In Eq. (\ref%
{amplhhg}), additionally to the above-mentioned assumptions, the further
approximation of considering only transitions from a bound state to a Volkov
state in the dipole moment has been made. In the single atom case, this has
been justified by the fact that the remaining contributions, from the
so-called \textquotedblleft continuum-to-continuum" transitions, were very
small. For a discussion of the various formulations of the SFA see, e.g.,
\cite{BLM94,hhgsfa,atisfa} and in particular \cite{BeckLew97}.

Due to the fact that it can be carried out almost entirely analytically, the
SFA is a very powerful approach. It possesses, however, the main drawback of
being gauge-dependent (for general discussions see, e.g., Ref. \cite{FKS96},
and for the specific case of molecules, Ref. \cite{PRACL2006}). Apart from
the obvious fact that the interaction Hamiltonians $H_{\mathrm{int}%
}(t^{\prime })$, which are present in $d_{\mathrm{ion}}(\mathbf{\tilde{p}}%
(t^{\prime }))$, are different in the length and velocity gauges \cite%
{footngauge}, in most computations where the SFA is employed, \textit{%
field-free} bound states are taken, which are not gauge equivalent. Indeed,
a field-free bound state $\left\vert \psi _{0}^{(L)}\right\rangle $ in the
length gauge would be gauge equivalent to the field-dressed state $%
\left\vert \psi _{0}^{(V)}\right\rangle =\chi _{v\leftarrow l}\left\vert
\psi _{0}^{(L)}\right\rangle $ in the velocity gauge, with $\chi
_{v\leftarrow l}=\exp [i\mathbf{A}(t)\cdot \mathbf{r}].$ Such a phase shift
causes a translation $\mathbf{p\rightarrow p-A}(t)$ on a momentum eigenstate
$\left\vert \mathbf{p}\right\rangle .$ Hence, for field-free bound states in
both gauges it leads to different dipole matrix elements $d_{\mathrm{rec}}(%
\mathbf{\tilde{p}}(t))$ and $d_{\mathrm{ion}}(\mathbf{\tilde{p}}(t))$.
Explicitly, in the length gauge $\mathbf{\tilde{p}}(t)=\mathbf{p}+\mathbf{A}%
(t)$, while in the velocity gauge $\mathbf{\tilde{p}}(t)=\mathbf{p}.$

For computations involving a single atom, the latter artifact can be avoided
by placing the system at the origin of the coordinate system. For systems
composed of several centers, such as molecules, however, this ambiguity will
always be present. Indeed, in the literature, different results have been
reported for molecular SFA computations in the velocity and in the length
gauge \cite{PRACL2006,DM2006,Madsen,Usachenko}. In recent, modified versions
of the length-gauge SFA, this problem has been eliminated for ATI by
considering the initial bound state $\left\vert \tilde{\psi}%
_{0}^{(L)}\right\rangle =\exp [-i\mathbf{A}(t^{\prime })\cdot \mathbf{r}%
]\left\vert \psi _{0}^{(L)}\right\rangle .$ Such a state is gauge-equivalent
to a field-free bound state in the velocity gauge, and, physically, may be
interpreted as a field-dressed state, in which the laser-field polarization
is included \cite{dressedSFA,DM2006}. For HHG, one may proceed in a similar
way, with the main difference that the dressing should also be included in
the final state, with which the electron recombines. In the dressed\
modified SFA, $\mathbf{\tilde{p}}(t)=\mathbf{p}$.

\subsection{Diatomic molecules}

\label{diatomic}

We will now apply the SFA to a diatomic molecule. For this purpose, we will
consider the simplest scenario, namely a one-electron system and frozen
nuclei. Furthermore, we will assume that the molecular orbital from which
the electron is released and with which it may recombine is a linear
combination of atomic orbitals (LCAO approximation). Explicitly, the
molecular bound-state wave function reads
\begin{equation}
\psi _{0}(r)=C_{\psi }(\psi _{0}^{(1)}(\mathbf{r}_{1})+\epsilon \psi
_{0}^{(2)}(\mathbf{r}_{2})),
\end{equation}%
where $\epsilon =\pm 1,\mathbf{r}_{1}=\mathbf{r}-\mathbf{R}/2,$ and $\mathbf{%
r}_{2}=\mathbf{r}+\mathbf{R}/2$ denote the positions of the centers $C_{1}$
and $C_{2},$ respectively, and $C_{\psi }$ is a normalization constant. \
For homonuclear molecules, which we will consider here, $\psi
_{0}^{(1)}=\psi _{0}^{(2)}=\varphi _{0}$. The positive and negative signs
for $\epsilon $ correspond to bonding and antibonding orbitals,
respectively. Within this context, there exist two main approaches for
computing high-order harmonic spectra, which will be discussed next.

\subsubsection{Prefactors}

The simplest and most widely used \cite%
{MBBF00,Madsen,Usachenko,moreCL,KB2005,DM2006,JMOCL2006} approach is to
incorporate the structure of the molecules in the prefactors $d_{\mathrm{rec}%
}(\mathbf{\tilde{p}}(t))$ and $d_{\mathrm{ion}}(\mathbf{\tilde{p}}(t)),$ and
to employ the same action $S(t,t^{\prime },\Omega ,\mathbf{p})$ as in the
single-atom case. The multiple integral in the transition amplitude (\ref%
{amplhhg}) can then either be solved numerically, or using saddle point
methods \cite{orbitshhg}. The latter procedure can be applied for high
enough intensities and low enough frequencies, and consists in approximating
(\ref{amplhhg}) by its asymptotic expansion around the coordinates $%
(t_{s},t_{s}^{\prime },\mathbf{p}_{s})$ for which $S(t,t^{\prime },\Omega ,%
\mathbf{p})$ is stationary. This implies that $\partial _{t}S(t,t^{\prime
},\Omega ,\mathbf{p})=\partial _{t^{\prime }}S(t,t^{\prime },\Omega ,\mathbf{%
p})=0$ and $\partial _{\mathbf{p}}S(t,t^{\prime },\Omega ,\mathbf{p})=%
\mathbf{0.}$ In this paper, we employ the specific saddle-point
approximations discussed in Ref. \cite{atiuni}.

For a single atom placed at the origin of the coordinate system, this leads
to the equations
\begin{equation}
\left[ \mathbf{p}+\mathbf{A}(t^{\prime })\right] ^{2}=-2I_{p},
\label{saddle1}
\end{equation}%
\begin{equation}
\int_{t^{\prime }}^{t}d\tau \left[ \mathbf{p}+\mathbf{A}(\tau )\right] =0,
\label{saddle3}
\end{equation}%
and
\begin{equation}
2(\Omega -I_{p})=\left[ \mathbf{p}+\mathbf{A}(t)\right] ^{2}.
\label{saddle2}
\end{equation}%
Eq. (\ref{saddle1}) expresses the conservation of energy at the time $\
t^{\prime }$ at which the electron reaches the continuum by tunneling
ionization. This equation possesses no real solution, which reflects the
fact that tunneling has no classical counterpart. In the limit $%
I_{p}\rightarrow 0$, one obtains such a condition for a classical particle
reaching the continuum with vanishing drift velocity. Eq. (\ref{saddle3})
constrains the intermediate momentum of the electron, so that it returns to
its parent ion, and, finally, Eq. (\ref{saddle2}) describes the conservation
of energy at a later time $t$, when the electron recombines with its parent
ion and a high-frequency photon of frequency $\Omega $ is generated.

The matrix element $d_{\mathrm{rec}}(\mathbf{\tilde{p}})=\left\langle
\mathbf{\tilde{p}}\right\vert \mathcal{O}_{\mathrm{dip}}\cdot \mathbf{e}%
_{x}\left\vert \psi _{0}\right\rangle $ then reads
\begin{equation}
d_{\mathrm{rec}}(\mathbf{\tilde{p}})=\frac{C_{\psi }}{(2\pi )^{3/2}}\left[
e^{i\mathbf{\tilde{p}\cdot R}/2}\mathcal{I}(\mathbf{r}_{1})+\epsilon e^{-i%
\mathbf{\tilde{p}\cdot R}/2}\mathcal{I}(\mathbf{r}_{2})\right] ,
\label{dipLCAO}
\end{equation}%
where
\begin{equation}
\mathcal{I}(\mathbf{r}_{j})=\int \mathcal{O}_{\mathrm{dip}}(\frac{\mathbf{r}%
_{1}+\mathbf{r}_{2}}{2})\cdot \mathbf{e}_{x}\exp [i\mathbf{\tilde{p}\cdot r}%
_{j}]\varphi _{0}(\mathbf{r}_{j})d^{3}r_{j},  \label{prefint1}
\end{equation}%
and $\mathcal{O}_{\mathrm{dip}}(\frac{\mathbf{r}_{1}+\mathbf{r}_{2}}{2})$ is
the dipole moment. In the length gauge, which we are mostly adopting in this
paper, $\ \mathbf{\tilde{p}}(t)=\mathbf{p}+\mathbf{A}(t).$ Unless strictly
necessary, in order to simplify the notation, we do not include the time
dependence in $\mathbf{\tilde{p}}.$ The dipole moment can be written in
several forms. If one considers the length form a natural choice is $%
\mathcal{O}_{\mathrm{dip}}^{(l)}(\mathbf{r})=-e\mathbf{r}+e\mathbf{r}_{1}+e%
\mathbf{r}_{2}=e\mathbf{r.}$ Other possibilities are to consider the
operator $\mathcal{O}_{dip}$ in its velocity and acceleration forms \cite%
{JMOCL2006,Ehrenfest}.

Inserting $\mathcal{O}_{\mathrm{dip}}(\mathbf{r})$ in Eq. (\ref{prefint1})
yields
\begin{equation}
\mathcal{I}(\mathbf{r}_{j})\propto \mathcal{I}_{j}(\mathbf{r}_{j})+\mathcal{I%
}_{j\nu }(\mathbf{r}_{j}),\ \text{\ with }j=1,2\text{ and }\nu \neq j,
\end{equation}%
where
\begin{equation}
\mathcal{I}_{j}(\mathbf{r}_{j})=\int \mathcal{O}_{\mathrm{dip}}(r_{j})\cdot
\mathbf{e}_{x}\exp [i\mathbf{\tilde{p}}\cdot \mathbf{r}_{j}]\varphi _{0}(%
\mathbf{r}_{j})d^{3}r_{j}
\end{equation}%
and
\begin{equation}
\mathcal{I}_{j\nu }(\mathbf{r}_{j})=\int \mathcal{O}_{\mathrm{dip}}(r_{\nu
})\cdot \mathbf{e}_{x}\exp [i\mathbf{\tilde{p}}\cdot \mathbf{r}_{j}]\varphi
_{0}(\mathbf{r}_{j})d^{3}r_{j},\nu \neq j.\   \label{inte2}
\end{equation}%
Specifically, if the dipole is in the length form, the above-stated
integrals read
\begin{equation}
\mathcal{I}_{j}(\mathbf{r}_{j})=-\frac{i}{2}\partial _{\tilde{p}_{x}}\phi (%
\mathbf{\tilde{p}})
\end{equation}%
and
\begin{equation}
\mathcal{I}_{j\nu }(\mathbf{r}_{j})=\frac{1}{2}(-i\partial _{p_{x}}\phi (%
\mathbf{\tilde{p}})+\epsilon _{j}R_{x}\phi (\mathbf{\tilde{p}})),
\label{crossed}
\end{equation}%
where
\begin{equation}
\phi (\mathbf{\tilde{p}})=\int \exp [i\mathbf{\tilde{p}\cdot r}_{j}]\varphi
_{0}(\mathbf{r}_{j})d^{3}r_{j}.  \label{pspacephi}
\end{equation}

In $\mathcal{I}_{j\nu }$ , $\epsilon _{j}=+1$ for $j=1$ and \ $\epsilon
_{j}=-1$ for $j=2$. Eq. (\ref{dipLCAO}) is then explicitly written as
\begin{equation}
d_{\mathrm{rec}}^{(b)}(\mathbf{\tilde{p}})=\frac{2iC_{\psi }}{(2\pi )^{3/2}}%
\left[ -\cos (\vartheta )\partial _{p_{x}}\phi (\mathbf{\tilde{p}})+\frac{%
R_{x}}{2}\sin (\vartheta )\phi (\mathbf{\tilde{p}})\right] ,  \label{prefb}
\end{equation}%
for bonding molecular orbitals (i.e., $\epsilon >0),$ or
\begin{equation}
d_{\mathrm{rec}}^{(a)}(\mathbf{\tilde{p}})=\frac{2C_{\psi }}{(2\pi )^{3/2}}%
\left[ \sin (\vartheta )\partial _{p_{x}}\phi (\mathbf{\tilde{p}})-\frac{%
R_{x}}{2}\cos (\vartheta )\phi (\mathbf{\tilde{p}})\right] ,  \label{prefa}
\end{equation}%
in the antibonding case (i.e., $\epsilon <0),$ with $\vartheta =\mathbf{%
\tilde{p}}\cdot \mathbf{R}/2.$

In Eqs. (\ref{prefb}) and (\ref{prefa}), the terms with a purely
trigonometric dependence on the internuclear distance yield the double-slit
condition in \cite{doubleslit}. The maxima and minima in the spectra which
are caused by this condition are expected to occur for
\begin{equation}
\mathbf{\tilde{p}}\cdot \mathbf{R}=2n\pi \text{ and }\mathbf{\tilde{p}}\cdot
\mathbf{R}=(2n+1)\pi ,  \label{maxmin}
\end{equation}%
respectively, for bonding molecular orbitals (i.e., $\epsilon >0).$ For
antibonding orbitals, the conditions are reversed, i.e., the maxima occur
for the odd multiples of $\pi $ and the minima for the even multiples. If
the velocity gauge is taken, the above stated conditions hold for $\mathbf{%
\tilde{p}}(t)=\mathbf{p}$, instead of $\ \mathbf{\tilde{p}}(t)=\mathbf{p}+%
\mathbf{A}(t)$. This is due to the fact that the initial and final
free-field sttes are not gauge equivalent, as discussed in Sec. IIA.

The remaining terms grow linearly with the projection $R_{x}$ of the
internuclear distance along the direction of the laser-field polarization,
and may lead to unphysical results \cite{PRACL2006,JMOCL2006}. For this
reason, they are sometimes neglected in the integrals $\mathcal{I}_{j\nu }(%
\mathbf{r}_{j})$ \cite{Kansas}. There exists, however, no rigorous
justification for such a procedure. Indeed, only recently, it has been shown
that such terms can be eliminated by considering an additional interaction
which depends on the nuclear coordinate. This interaction is present in a
modified molecular SFA, in its dressed and undressed versions \cite{DM2006},
and leads to contributions which cancel out the linear term in $R_{x}$.

In the length gauge, if the length form of $\mathcal{O}_{\mathrm{dip}}$ is
taken, $d_{\mathrm{rec}}(\mathbf{\tilde{p}}(t))=d_{\mathrm{ion}}(\mathbf{%
\tilde{p}}(t^{\prime }))$, with $\ \mathbf{\tilde{p}}(t)=\mathbf{p}+\mathbf{A%
}(t),$ while in the velocity gauge,
\begin{equation}
d_{\mathrm{ion}}^{(b)}(\mathbf{\tilde{p}})=\frac{C_{\psi
}[\mathbf{p}+\mathbf{A}(t^{\prime })]^{2}}{(2\pi )^{3/2}}\cos
(\vartheta )\phi (\mathbf{\tilde{p}}),
\end{equation}%
or%
\begin{equation}
d_{\mathrm{ion}}^{(a)}(\mathbf{\tilde{p}})=-i\frac{C_{\psi
}[\mathbf{p}+\mathbf{A}(t^{\prime })]^{2}}{(2\pi )^{3/2}}\sin
(\vartheta )\phi (\mathbf{\tilde{p}}),
\end{equation}%
with $\mathbf{\tilde{p}}(t)=\mathbf{p,}$ for bonding and antibonding
molecular orbitals, respectively.

\subsubsection{Modified saddle-point equations}

Physically, if one employs the pre-factors (\ref{prefb}) and (\ref{prefa}),
this means that one is not modifying the saddle-point equations (\ref%
{saddle1})-(\ref{saddle2}). Therefore, the orbits along which the electron
is moving in the continuum remain the same as in the single-center case.
This approach is questionable in several ways. From the technical viewpoint,
there is no guarantee that such pre-factors are slowly varying, as compared
to the semiclassical action, especially for large internuclear distances
\cite{HenrikDipl,PRACL2006}. Furthermore, in general, they do not
incorporate processes in which the electron leaves one center of the
molecule and recombines with the other, which, physically, are expected to
be present in molecular HHG \cite{PRACL2006}.

A slightly more sophisticated approach is to exponentialize the pre-factors
obtained in the two-center case and incorporate the terms in $\mathbf{\tilde{%
p}\cdot R}/2$ in the action. This procedure has been adopted in \cite%
{PRACL2006} and will be closely followed in this work. For the sake of
simplicity, we will consider the modified prefactor
\begin{equation}
\tilde{d}_{\mathrm{rec}}^{(b)}(\mathbf{\tilde{p}})=-\frac{2iC_{\psi }}{(2\pi
)^{3/2}}\left[ \cos \left( \mathbf{\tilde{p}}\cdot \frac{\mathbf{R}}{2}%
\right) \partial _{\tilde{p}_{x}}\phi (\mathbf{\tilde{p}})\right] ,
\label{modifieddip}
\end{equation}%
for which the second term in Eq. (\ref{crossed}) is absent. Eq. (\ref%
{modifieddip}) is also very similar to the dipole matrix element in the
velocity form, apart from the fact that, in the latter case, $\partial _{%
\tilde{p}_{x}}\phi (\mathbf{\tilde{p}})$ is replaced by $\mathbf{\tilde{p}}%
\phi (\mathbf{\tilde{p}})$ \cite{JMOCL2006}. In the expression for the
antibonding case, the cosine term in (\ref{modifieddip}) should be replaced
by $\sin (\mathbf{\tilde{p}\cdot R}/2)$.

This leads to the sum
\begin{equation}
M=\sum_{j=1}^{2}\sum_{\nu =1}^{2}M_{j\nu }  \label{sumampl}
\end{equation}%
of the transition amplitudes
\begin{eqnarray}
M_{j\nu } &=&\frac{C_{\psi }}{(2\pi )^{3/2}}\int_{0}^{t}dt^{\prime }\int
dt\int d^{3}p\eta (\mathbf{p},t,t^{\prime })  \notag \\
&&\times \exp [iS_{j\nu }(\mathbf{p},\Omega ,t,t^{\prime })],\
\label{amplitudes}
\end{eqnarray}%
with $\eta (\mathbf{p},t,t^{\prime })=\left[ \partial _{\tilde{p}_{x}}\phi (%
\mathbf{\tilde{p}}(t))\right] ^{\ast }\partial _{\tilde{p}_{x}}\phi (\mathbf{%
\tilde{p}(}t^{\prime })).$ The terms $S_{j\nu }$ correspond to a modified
action, which incorporates the structure of the molecule.

Explicitly, for the undressed length-gauge SFA,
\begin{equation}
S_{jj}=S(\mathbf{p},\Omega ,t,t^{\prime })+(-1)^{j+1}\xi _{1}(R,t,t^{\prime
})  \label{ssame}
\end{equation}%
and
\begin{equation}
S_{j\nu }=S(\mathbf{p},\Omega ,t,t^{\prime })+(-1)^{\nu +1}\xi
_{2}(R,t,t^{\prime }),\ \nu \neq j,  \label{sdiff}
\end{equation}%
where $\xi _{1}(R,t,t^{\prime })=[\mathbf{A}(t)-\mathbf{A}(t^{\prime
})]\cdot \mathbf{R}/2$ and $\xi _{2}(R,t,t^{\prime })=\mathbf{p}\cdot
\mathbf{R+}[\mathbf{A}(t)+\mathbf{A}(t^{\prime })]\cdot \mathbf{R}/2.$ Eq. (%
\ref{ssame}) and (\ref{sdiff}) may be directly related to physical processes
involving one or two centers in the molecule, respectively, as it will be
discussed next.

For this purpose, we will solve the multiple integrals in (\ref{amplitudes})
employing saddle-point methods. The conditions $\partial _{\mathbf{p}%
}S_{j\nu }(\mathbf{p},\Omega ,t,t^{\prime })=\mathbf{0},\ \partial
_{t}S_{j\nu }(\mathbf{p},\Omega ,t,t^{\prime })=0$ and $\partial _{t^{\prime
}}S_{j\nu }(\mathbf{p},\Omega ,t,t^{\prime })=0$ upon the derivative of the
action yield saddle-point equations, which, as in the previous section, can
be related to the orbits of an electron recombining with the molecule.
Explicitly, for the modified action $S_{jj},$ [Eq. (\ref{ssame})], the
saddle-point equations read
\begin{equation}
\frac{\lbrack \mathbf{p}+\mathbf{A}(t^{\prime })]^{2}}{2}=-I_{p}-\mathbf{E}%
(t^{\prime })\cdot \mathbf{R}/2,  \label{tunnelC1}
\end{equation}%
\begin{equation}
\frac{\lbrack \mathbf{p}+\mathbf{A}(t)]^{2}}{2}=\Omega -I_{p}+\mathbf{E}%
(t)\cdot \mathbf{R}/2,  \label{recC1}
\end{equation}%
for $\ j=1,$ or
\begin{equation}
\frac{\lbrack \mathbf{p}+\mathbf{A}(t^{\prime })]^{2}}{2}=-I_{p}+\mathbf{E}%
(t^{\prime })\cdot \mathbf{R}/2,  \label{tunnelC2}
\end{equation}%
\begin{equation}
\frac{\lbrack \mathbf{p}+\mathbf{A}(t)]^{2}}{2}=\Omega -I_{p}-\mathbf{E}%
(t)\cdot \mathbf{R}/2,  \label{recC2}
\end{equation}%
for $\ j=2.$ The saddle point equations (\ref{tunnelC1}) and (\ref{tunnelC2}%
) correspond to the tunnel ionization process from center $C_{1}$ and $C_{2}$%
, respectively. Curiously, both equations contain extra terms, if compared
to equation (\ref{saddle1}) for the single-atom case. Such terms are
dependent on the internuclear distance $\mathbf{R}$ and the external laser
field $\mathbf{E}(t^{\prime })$ at the time the electron is freed, and may
be interpreted as potential-energy shifts in the barrier through which the
electron tunnels out. Similar terms are also observed in Eqs. (\ref{recC1})
and (\ref{recC2}) for the energy conservation at the time the electron
recombines, as compared to the single-atom expression (\ref{saddle2}).

The remaining saddle point equation is given by the same expression as in
the single-atom case, i.e., Eq. (\ref{saddle3}), and means that, for $M_{11}$
and $M_{22}$, the electron is ejected and returns to the same center. If on
the other hand, we consider the modified action (\ref{sdiff}), this yields
\begin{equation}
\int_{t^{\prime }}^{t}[\mathbf{p}+\mathbf{A}(s)]ds\pm \mathbf{R}=0,
\label{returndiff}
\end{equation}%
which, physically, mean that the electron is leaving from one center and
recombining with the other. The negative and positive signs refer to the
transition amplitudes $M_{12}$ (center $C_{1}$ to center $C_{2})$ and $%
M_{21} $ (center $C_{2}$ to center $C_{1}$), respectively. In the former
case, the remaining saddle points are given by (\ref{tunnelC1}) and (\ref%
{recC2}), i.e., the electron tunnels from $C_{1}$ and recombines with $%
C_{2}, $ whereas in the latter case they are given by (\ref{tunnelC2}) and (%
\ref{recC1}), which physically, expresses the fact that the electron is
ejected at $C_{2}$ and recombines with $C_{1}.$

The energy shifts $\pm \mathbf{E}(\tau )\cdot \mathbf{R}/2,$ $\tau
=t,t^{\prime }$ in (\ref{tunnelC1})-(\ref{recC2}) are absent in the velocity
gauge SFA \cite{PRACL2006}, and in a modified length-gauge SFA, in which the
electric field polarization is incorporated in the initial and final
electronic bound states. In both cases, the action $S_{jj}$, associated to
orbits involving only one center, is given by the single-atom expression (%
\ref{actionhhg}), which leads to the saddle-point equations (\ref{saddle1})-(%
\ref{saddle2}). The action $S_{j\nu }$ related to two-center orbits reads
\begin{equation}
\tilde{S}_{j\nu }=S(\mathbf{p},\Omega ,t,t^{\prime })+(-1)^{\nu +1}\mathbf{p}%
\cdot \mathbf{R},\ \nu \neq j.  \label{sdiffmod}
\end{equation}%
The above-stated expression leads to the single-atom equations (\ref{saddle1}%
), (\ref{saddle2}) for tunneling and rescattering, together with the
two-center return condition (\ref{returndiff}). The prefactors, however, are
different in both cases. In the dressed modified length-gauge SFA, $\eta
^{D}(\mathbf{p},t,t^{\prime })=\left[ \partial _{p_{x}}\phi (\mathbf{p})%
\right] ^{\ast }\partial _{p_{x}}\phi (\mathbf{p}),$ while in the velocity
gauge $\eta ^{V}(\mathbf{p},t,t^{\prime })=\left[ \partial _{p_{x}}\phi (%
\mathbf{p})\right] ^{\ast }\phi (\mathbf{p})[\mathbf{p}+\mathbf{A}(t^{\prime
})]^{2}/2.$

One should note that, within the specific model employed here, the physical
process in which the electron moves directly from one center to the other,
without reaching the continuum, is not being considered. Such a process
leads to a strong set of harmonics in the low-energy range of the spectra.
Since, however, we are focusing on the plateau harmonics, the contributions
from this extremely short set of orbits are not of interest to the present
discussion. For a detailed study of this case, see, e.g., \cite%
{KBK98,Usach2006}.

\subsubsection{Additional attosecond pulses}

\label{attopulses}

The role of the energy shifts observed in (\ref{tunnelC1})-(\ref{recC2})is
not well understood. A way of eliminating such terms is to modify the
length-gauge SFA and include the influence of the laser field in the initial
and final states. However, even without such modifications, it is possible
to provide an additional pathway for the electron to reach the continuum, so
that, at least in the context of tunneling ionization, these shifts can be
avoided. For instance, if the electron is ejected by a high-frequency
photon, it does not have to tunnel through potential barriers with energy
shifts whose physical meaning is not clear. Such a pathway can be provided
by a time-delayed attosecond-pulse train $\mathbf{E}_{\mathrm{h}}(t)$
superposed to a strong, near infra-red field $\mathbf{E}_{l}(t)=E_{0}\sin
\omega t\mathbf{e}_{x}$. Indeed, it has been recently shown that such pulses
can be used to control the electron ejection in the continuum, and thus
high-harmonic generation and above-threshold ionization \cite%
{atthhg2004,attati2005,FSVL2006}.

In \cite{FSVL2006,FS2006}, we employed such a scheme to control
quantum-interference effects for high-harmonic generation and
above-threshold ionization for the single-atom case, within the SFA
framework. Our previous findings suggest that the probability of the
electron reaching the continuum, in case it is ejected by the attosecond
pulses, is roughly the same for all sets of orbits. Indeed, it appears that
the sole, or at least main factor determining the intensities in the spectra
is the excursion time of the electron in the continuum. In the specific case
studied in \cite{FSVL2006}, there was a set of very short orbits, which led
to particularly strong harmonics. Therefore, an attosecond pulse train
superposed to a strong laser field is an ideal setup to avoid any artifacts
due to modified tunneling conditions.

In \cite{FSVL2006}, we have approximated the attosecond pulse train by a sum
of Dirac-Delta functions in the time domain. This yields
\begin{equation}
\mathbf{E}_{\mathrm{h}}(t)=E_{h}\pi \sum_{n=0}^{\infty }\frac{(-1)^{n}}{%
\sigma (t)}\delta (t-\frac{n\pi }{\omega }-t_{d})\mathbf{\epsilon }_{x},
\label{attotrain}
\end{equation}%
where $\omega $, $E_{h}$, and $\sigma (t)$ denote the laser field frequency,
the attosecond-pulse strength and the train temporal envelope, respectively.
This approximation is the limiting case for a train composed of an infinite
set of harmonics, and has the main advantage of allowing an analytic
treatment of the transition amplitudes involved up to at most one numerical
integration. Furthermore, it is a reasonable asymptotic limit for pulses
composed by a large high-harmonic set. We consider here $\sigma =const,$
which, physically, corresponds to an infinitely long attosecond-pulse train.
Clearly, the total field is given by $\mathbf{E}(t)=\mathbf{E}_{l}(t)+%
\mathbf{E}_{\mathrm{h}}(t).$

We will now assume that the attosecond pulses are the only cause of
ionization and that the subsequent propagation of the electron in the
continuum is determined only by the monochromatic field. Hence, $\mathbf{E}%
(t^{\prime })\simeq \mathbf{E}_{\mathrm{h}}(t^{\prime })$ and $\mathbf{A}%
(t)\simeq \mathbf{A}_{l}(t)$ in Eqs.~(\ref{amplhhg}) and (\ref{amplitudes}).
This eliminates the integral over the ionization time in the transition
amplitudes. Hence, the values for which the single atom-action $%
S(t,t^{\prime },\Omega ,\mathbf{k}),$ or the modified action $S_{j\nu }$ is
stationary must be determined only with respect to the variables $t$ and $%
\mathbf{p}$. Physically this means that the recombination and return
conditions remain the same, with regard to the purely monochromatic case,
but that there is no longer a saddle-point equation constraining the initial
electron momentum. In fact, the electron is being ejected in the continuum
with any of the energies $N\omega -I_{p}$, since all harmonics composing the
train are equivalent. For the other extreme limit, namely a high-frequency
monochromatic wave, we refer to \cite{FS2006}, where we provide a detailed
discussion within the SFA.

Explicitly, if the action is not modified, these assumptions lead to the
transition amplitude
\begin{eqnarray}
M_{\mathrm{h}}^{(D)} &=&\frac{i\pi C_{\psi }E_{\mathrm{h}}}{\sigma (2\pi
)^{3/2}}\sum_{n=0}^{\infty }(-1)^{n}\int_{-\infty }^{+\infty }\hspace*{-0.5cm%
}dt\int d^{3}p\exp \left[ iS(\mathbf{p},\Omega ,t,t^{\prime })\right]  \notag
\\
&&\times d_{\mathrm{rec}}^{\ast }(\mathbf{\tilde{p}}(t))d_{\mathrm{ion}}(%
\mathbf{\tilde{p}}(t^{\prime })).  \label{hhgatto}
\end{eqnarray}%
In case one considers the transition amplitudes (\ref{amplitudes}), this
yields%
\begin{eqnarray}
M_{j\nu } &=&\frac{i\pi C_{\psi }E_{\mathrm{h}}}{\sigma (2\pi )^{3/2}}%
\sum_{n=0}^{\infty }(-1)^{n}\hspace*{-0.2cm}\int_{-\infty }^{+\infty }%
\hspace*{-0.2cm}\hspace*{-0.1cm}dt\int d^{3}p\eta (\mathbf{\tilde{p}}%
,t,t^{\prime }) \\
&&\times \exp [iS_{j\nu }(\mathbf{p},\Omega ,t,t^{\prime })].\
\end{eqnarray}%
In both equations, $t^{\prime }=t_{d}+n\pi /\omega .$ The saddle-point
equations (\ref{saddle3}) and (\ref{saddle2}) for the single atom case can
then be combined as \

\begin{eqnarray}
&&\sin \omega t-(-1)^{n}\sin \omega t_{d}  \notag \\
&=&\hspace*{-0.2cm}\left[ \omega (t-t_{d})-n\pi \right] \left( \cos \omega
t\mp \sqrt{\frac{\Omega -I_{p}}{2U_{p}}}\right) ,  \label{saddelt}
\end{eqnarray}%
which will give the return times $t.$ If, on the other hand, these equations
are modified, it is possible to distinguish four main scenarios.
Specifically, for the processes in which the electron leaves and returns to
the same center, the saddle-point equations differ from (\ref{saddelt}) only
in a shift $I_{p}\rightarrow I_{p}\pm \mathbf{E}_{l}(t)\cdot \mathbf{R}/2.$
The negative and positive signs correspond to $M_{11}$ and $M_{22},$
respectively. For the scenarios involving two centers, the saddle-point
equations read
\begin{eqnarray}
&&\sin \omega t-(-1)^{n}\sin \omega t_{d}+\epsilon \frac{R_{x}\omega }{2U_{p}%
}  \notag  \label{moddelt} \\
&=&\hspace*{-0.2cm}\left[ \omega (t-t_{d})-n\pi \right] \left( \cos \omega
t\mp \sqrt{\frac{\Omega -\tilde{I}_{p}}{2U_{p}}}\right) ,
\end{eqnarray}%
with $\tilde{I}_{p}=I_{p}+\epsilon \mathbf{E}_{l}(t)\cdot \mathbf{R}/2.$ The
case $\epsilon =-1$ and $\epsilon =+1$ corresponds to $M_{12}$ and $M_{21},$
respectively.

\section{Harmonic spectra}

\label{results} We will now present high-harmonic spectra, in the presence
and absence of the attosecond pulses. We restrict the electron ionization
times to the first half cycle of the driving field. We also consider the six
shortest pairs of orbits for the returning electron. Due to wave-packet
spreading, the contributions from the remaining pairs are negligible \cite%
{spreading}. For simplicity, we employ a bonding combination of 1s states,
for which
\begin{equation}
\phi (\mathbf{\tilde{p}})\sim \frac{1}{[\mathbf{\tilde{p}}^{2}+2I_{p}]^{2}},
\label{dip1s}
\end{equation}%
and assume that the molecule is aligned parallel to the laser-field
polarization, so that $R_{x}=R.$

\subsection{Prefactors}

\label{singlecenter} 
\begin{figure}[tbp]
\begin{center}
\includegraphics[width=9cm]{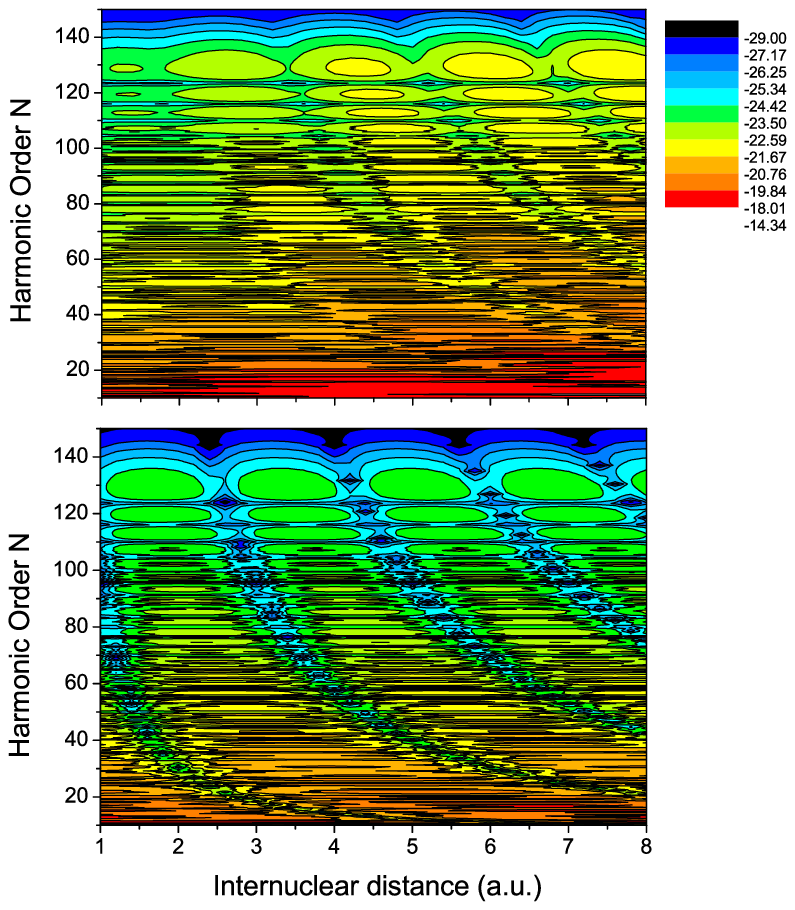}
\end{center}
\caption{(Color Online) High-harmonic spectra computed employing the
single-atom orbits and two center prefactors, using a bonding ($\protect%
\epsilon =+1$), linear combination of atomic orbitals, for internuclear
distances $1$ a.u.$\leq R\leq 8$ a.u. The upper and lower panels correspond
to the dipole prefactors ({\protect\ref{prefb}}) and ({\protect\ref%
{modifieddip}}), respectively. The atomic system was chosen as $H_{2}^{+}$,
which was approximated by the linear combination of $1s$ atomic orbitals
with $I_{p}=0.5$ a.u.. We took the the driving field intensity and frequency
to be $I=1\times 10^{15}\mathrm{W/cm^{2}}$, and $\protect\omega =0.057$
a.u., respectively. }
\label{contour}
\end{figure}

We will commence by considering the single-center
action (\ref{actionhhg}) and the prefactors discussed in the previous
section. In Fig. \ref{contour}, we depict high-harmonic spectra computed for
a wide range of internuclear distances, employing the prefactor (\ref{prefb}%
) or the modified expression (\ref{modifieddip}), for which the linear term
in $R_{x}$ is absent [upper and lower panel, respectively]. The figure
illustrates how the linear term masks the interference patterns. In fact,
for a.u. $1\leq R\leq 3$ a.u., it counterbalances the influence of the
purely trigonometric term, and no clear minima and maxima are observed. As
the internuclear distance increases, this term starts to play a dominant
role, displacing the maxima and minima away from the double-slit condition (%
\ref{maxmin}). Furthermore, it seems that the patterns become more distinct
with increasing harmonic order.

\begin{figure}[tbp]
\begin{center}
\noindent\includegraphics[width=9cm]{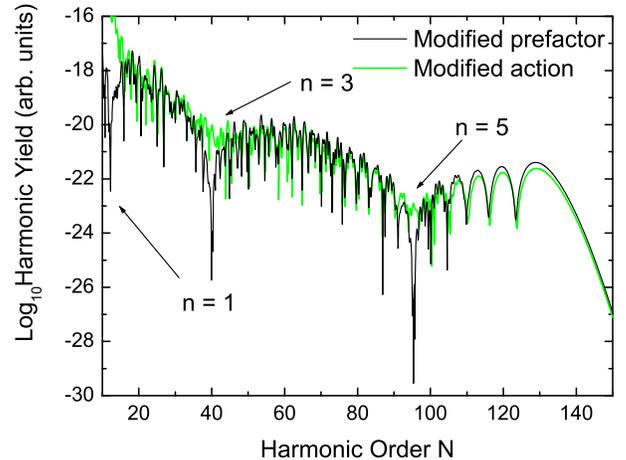}
\end{center}
\caption{(Color online) Spectra computed employing the single-atom orbits
and two center prefactors (black thin lines), as compared to those obtained
employing modified saddle-point equations (green thick lines). We consider
here the modified length form ({\protect\ref{modifieddip}}) of the dipole
operator, which excludes the term with a linear dependence on $R_{x}$. The
atomic system was chosen as $H_{2}^{+}$, which was approximated by the
linear combination of $1s$ atomic orbitals with $I_{p}=0.5$ a.u.. The
internuclear distance and the alignment angle are $R=5$ a.u., and $\protect%
\theta =0,$ respectively. The driving field intensity and frequency are
given by $I=1\times 10^{15}\mathrm{W/cm^{2}}$, and $\protect\omega =0.057$
a.u., respectively. The interference minima are indicated by the arrows in
the figure.}
\label{interfe1}
\end{figure}

A rough estimate of the influence of each term in (\ref{prefb}) on the
spectra agrees with Fig.~1. Since the trigonometric functions in (\ref{prefb}%
) are bounded, the ratio between the maxima caused by each term will be \ $%
\varsigma =R|\cos \theta \phi (\mathbf{\tilde{p}})/(2\partial _{p_{x}}\phi (%
\mathbf{\tilde{p}}))|$, where $\theta $ is the alignment angle and $\phi (%
\mathbf{\tilde{p}})$ is given by Eq. (\ref{pspacephi}). If $\varsigma \simeq
1$, the maxima will possess the same order of magnitude and there will be no
noticeable modulation, while if $\varsigma <1$ the double-slit physical
picture may still be reproduced. For $\varsigma >1,$ however, one expects
that the linear term in $R$ will prevail. Hence, the critical value for the
internuclear distance is $R_{c}=2|\sec \theta \partial _{p_{x}}\phi (\mathbf{%
\tilde{p}})/\phi (\mathbf{\tilde{p}})|$. This expression depends on the
bound states with which the electron recombines, and also on the harmonic
energy. For instance, specifically for $1s$ states, $R_{c}\sim 4{\tilde{p}%
_{x}}\sec \theta /(\mathbf{\tilde{p}}^{2}+2I_{p}).$ Above the ionization
threshold, according to Eq. (\ref{saddle2}), $R_{c}\sim 2\tilde{p}_{x}\sec
\theta /\Omega $. Hence, one expects the linear term to be more prominent as
the harmonic energy increases, leading to clearer, though incorrect,
patterns. In order to avoid such problems, we will employ the prefactor (\ref%
{modifieddip}) throughout.

In Fig. \ref{interfe1}, we compare spectra computed using either the
pre-factor (\ref{modifieddip}) or modified saddle-point equations. Both
spectra are very similar, with maxima and minima at harmonic frequencies $%
\Omega =I_{p}+n^{2}\pi ^{2}/(2R_{x}^{2})$, as expected from the double-slit
condition. This similarity holds not only for the gross features, but,
additionally, for the substructure caused by other types of quantum
interference. Close to the minima, however, the yield from the latter case
is larger. Nevertheless, the very good overall agreement shows that, in
fact, the patterns obtained can also be interpreted as the result of the
quantum interference between different types of electron orbits.

\subsection{Interference effects}

\label{orbits} 
\begin{figure}[tbp]
\begin{center}
\noindent\includegraphics[width=9cm]{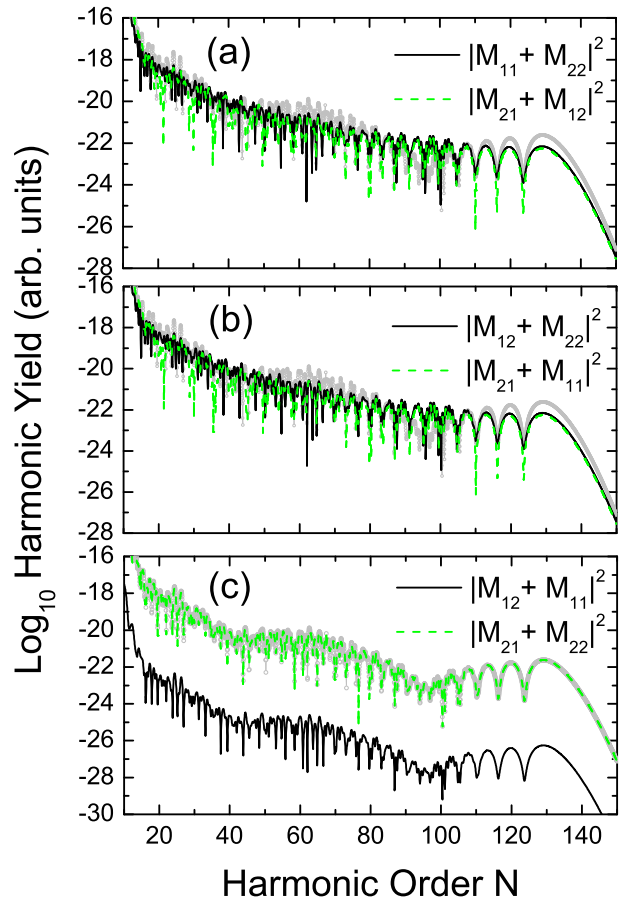}
\end{center}
\caption{(Color Online) Contributions to the high-harmonic yield from the
quantum interference between different types of orbits, for internuclear
distance $R=5$ a.u. The remaining parameters are the same as in Fig. 3.
Panel (a): Orbits involving similar scattering scenarios, i. e., $%
|M_{11}+M_{22}|^{2}$, and $|M_{12}+M_{21}|^{2}.$ Panel (b): Orbits \emph{%
ending} at the same center, i.e., $|M_{11}+M_{21}|^{2}$ and $%
|M_{12}+M_{22}|^{2}.$ Panel (e): Orbits \emph{starting} at the same center,
i.e., $|M_{11}+M_{12}|^{2}$ and $|M_{21}+M_{22}|^{2}.$ For comparison, the
full contributions $|M_{21}+M_{22}+M_{11}+M_{12}|^{2}$ are displayed as the
light gray circles in the picture.}
\label{orbits1}
\end{figure}

In order, however, to identify which sets of orbits cause the
dips and the maxima, we will analyze the interference between their
individual contributions. Such results are shown in Fig. \ref{orbits1}. In
the upper panels, we present the spectra computed from topologically similar
sets of orbits, i.e., from processes involving only one, or two centers
[panel \ref{orbits1}.(a)]. In this case, the main interference patterns are
absent. This strongly suggests that they are due to the quantum interference
of topologically different sets of orbits: the orbits along which an
electron leaves and returns to the same center, and those along which it
reaches the continuum at one center and recombines with the other.
Physically, this could be attributed to the fact that, in this case, there
would be an appreciable phase difference between the two sets of orbits,
since the latter are much longer than the former. This phase difference
would cause the overall modulation.

Hence, there are two remaining possibilities. Concretely, the modulation can
be due to the quantum interference either between processes in which the
electron leaves from different centers and recombines with the same center
(i.e., between the orbits which start at $C_{j}$ and end at $C_{\nu }$, with
$\nu \neq j,$ and those starting and ending at $C_{\nu }$, and $j=1,2$), or
between those in which an electron starts at the same center and recombines
with different centers (i.e., the electron is ejected at $C_{j}$ and
recombines at $C_{\nu }$, $\nu \neq j,$ or it is freed and recombines at $%
C_{j}$, with $j=1,2$). In panel \ref{orbits1}.(b), we consider the former
processes, whereas in panel \ref{orbits1}.(c) we depict the latter.
Interestingly, only in case the electron leaves from the same center, the
interference patterns are present. Furthermore, there is a difference in
roughly four orders of magnitude between the two types of contributions.
Such a difference is absent in the other cases. Additionally, the full
contributions to the yield are practically indistinguishable from the
transition probability $|M_{21}+M_{22}|^{2},$ from the orbits starting at $%
C_{2}.$

\begin{figure}[tbp]
\begin{center}
\noindent\includegraphics[width=9cm]{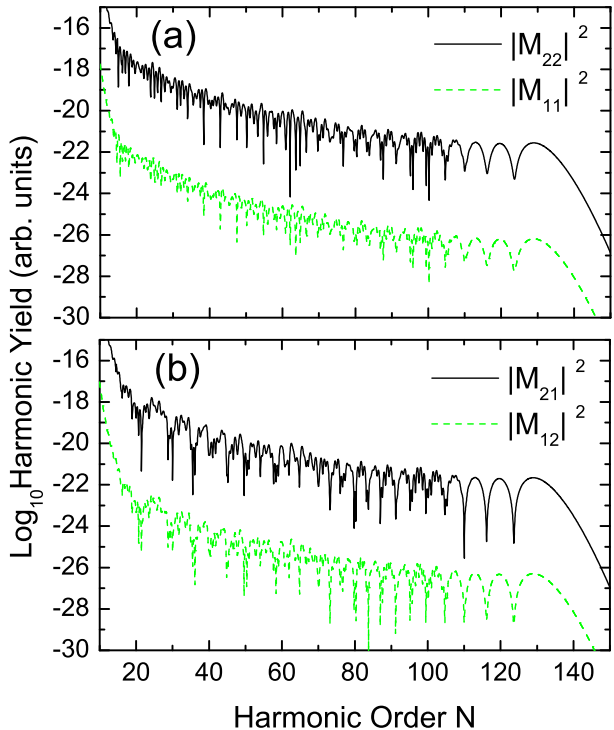}
\end{center}
\caption{(Color online) Contributions to the high-harmonic yield from
specific types of orbits, for internuclear distance $R=5$ a.u. The remaining
parameters are the same as in Fig. 3. Panel (a): Yield from the orbits
starting and ending at the same center, i. e., transition probabilities $%
|M_{11}|^{2}$ and $|M_{22}|^{2}.$ Panel (b): Yield from the orbits starting
and ending at different centers, i.e., transition probabilities $%
|M_{12}|^{2} $ and $|M_{21}|^{2}.$}
\label{orbits2}
\end{figure}

In order to understand this better, one must have a closer look at the
individual contributions from different sets of orbits, and, in particular
their orders of magnitude. Such contributions, depicted in Panels (a) and
(b) of Fig. \ref{orbits2}, show that the transition probabilities $%
|M_{22}|^{2}$ and $|M_{21}|^{2},$ which correspond to the orbits starting
from the center $C_{2},$ are roughly four orders of magnitude larger than $%
|M_{12}|^{2}$ and $|M_{11}|^{2}$, i.e., than those from the orbits starting
at $C_{1}$. Therefore, it is not surprising that the yield is dominated by $%
|M_{21}+M_{22}|^{2}$ in the previous figure. Furthermore, these results
exhibit no maxima and minima. Hence, they support the assumption that such
features are due to the interference of different types of orbits. Finally,
the contributions from orbits starting at the same center possess the same
order of magnitude. This suggests that tunneling ionization is the main
mechanism determining the relevance of a particular type of orbits to the
spectra, and that are local differences in the barrier through which the
electron must tunnel, depending on the center it starts from.

An inspection of the imaginary parts $\mathrm{Im}[t^{\prime }]$ of the start
times provides additional insight into this problem. Due to the fact that
tunneling ionization is a process which has no classical counterpart, this
quantity is always non-vanishing, even if the energy range in question is
lower than the maximal harmonic energy. The larger $\mathrm{Im}[t^{\prime }]$
is, the less probable it will be that tunneling ionization takes place. Such
an interpretation has been successfully employed in \cite{FLSL2004} in order
to determine the dominant pairs of orbits, in the context of nonsequential
double ionization with few-cycle laser pulses, and will be also considered
in this work. For that purpose, we will take the shortest pairs of orbits
utilized in the computation of the transition probabilities in Fig. \ref%
{orbits2} and, for each case, display $\mathrm{Im}[t^{\prime }]$. These are
the dominant pairs of orbits. The longer pairs have a less significant
influence on the spectra, due to the spreading of the electronic wave packet
\cite{spreading}.

\begin{figure}[tbp]
\begin{center}
\noindent\includegraphics[width=8cm]{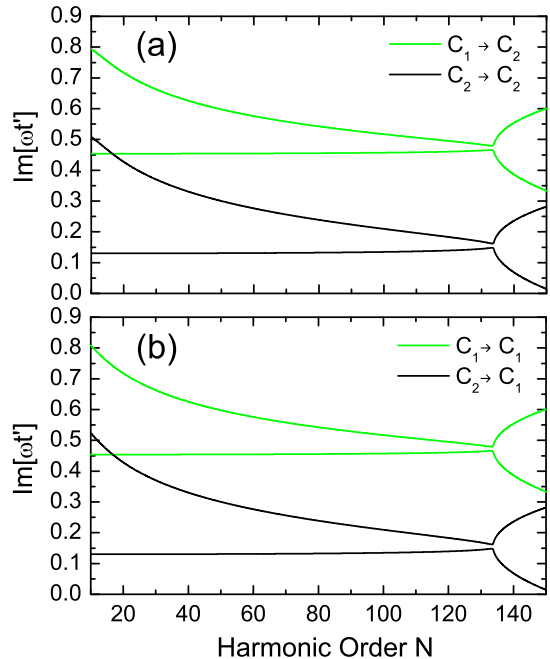}
\end{center}
\caption{(Color online) Imaginary parts of the start times $t^{\prime }$ of
the shortest pairs of orbits, which contribute to the matrix elements $%
M_{22},$ and $M_{12},$ [panel (a)], and to $M_{21},$ and $M_{11}$ [panel
(b)], for the same field and atomic parameters as in the previous figure. }
\label{orbits3}
\end{figure}

Such results are depicted in Fig. \ref{orbits3}. Clearly, $\mathrm{Im}%
[t^{\prime }]$ is around four times larger for the orbits starting from the
center $C_{1}$, as compared to those starting from $C_{2}.$ This means that,
in order to reach the continuum, the electron must overcome a larger barrier
if it comes from $C_{1}$. Since, roughly speaking, the ionization
probability per unit time decreases exponentially with $\mathrm{Im}%
[t^{\prime }]$, one expects the contributions from the orbits starting from $%
C_{1}$ to be around four orders of magnitude smaller than those from the
orbits starting at $C_{2}$. An inspection of Fig. \ref{orbits2} shows that
this is indeed the case.

The above-stated effect could, however, be an artifact of the strong-field
approximation in the length gauge. Indeed, the terms $\pm \mathbf{E}%
(t^{\prime })\cdot \mathbf{R}$ in the saddle-point equations (\ref{tunnelC1}%
) and (\ref{tunnelC2}) can be interpreted as potential energy shifts, due to
the fact that the electron is displaced from the origin \cite%
{PRACL2006,DM2006}. Such terms increase or sink the potential barrier for $%
C_{1}$ or $C_{2},$ respectively, and, consequently, change the orders of
magnitude in the contributions starting from different centers. Even though,
as a whole, the results match those obtained by other means, their physical
interpretation is controversial.

One may, however, avoid this problem by providing an additional pathway for
the electron to reach the continuum. For that purpose, we shall superpose a
time-delayed attosecond pulse train to the strong laser field, employing the
model discussed in Sec. \ref{attopulses} and in our previous work \cite%
{FSVL2006,FS2006}. The maximal harmonic energies for this specific model are
strongly dependent on the time delay $t_{d}$ between the attosecond-pulse
train the infra-red field, extending from the ionization potential, for $%
\phi =0.75\pi $ to $I_{p}+1.8U_{p}$, for $\phi =n\pi $. Furthermore, there
exist many intermediate delays, for which a double plateau is present. This
substructure may be detrimental to the identification and physical
interpretation of the interference patterns. In order to avoid such
problems, we will consider here vanishing time delay, i.e., $\phi =0.$
\begin{figure}[tbp]
\begin{center}
\includegraphics[width=8cm]{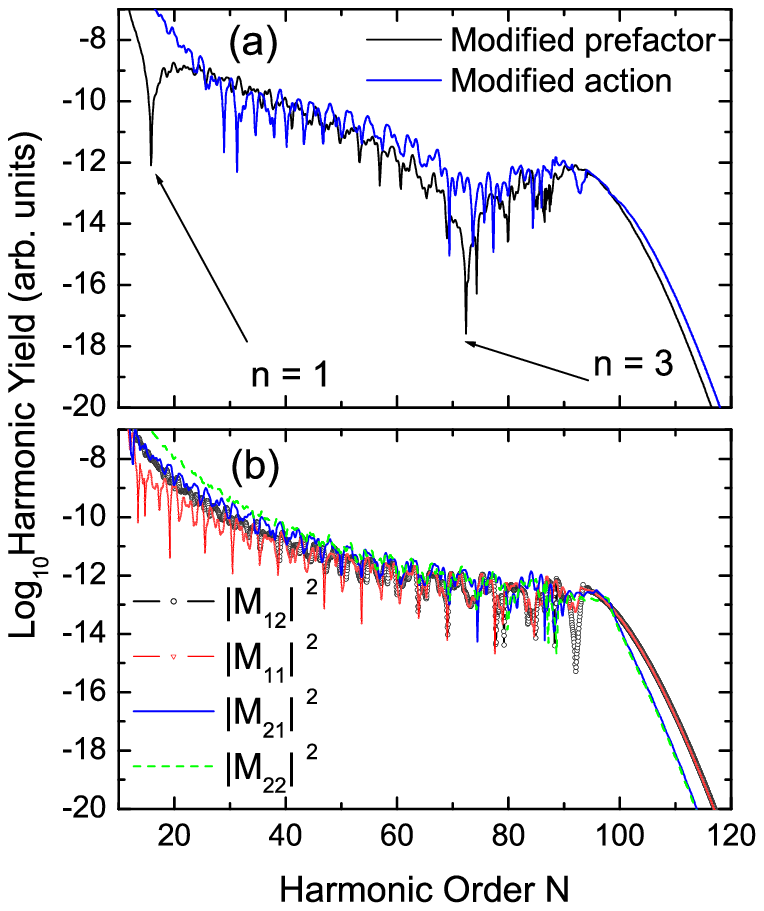}
\end{center}
\caption{(Color Online) High-harmonic spectra for a molecule aligned
parallel to the laser-field polarization, and internuclear distance $R=3.5$
a.u, for an attosecond pulse train superposed to a strong, near-infrared
field, and exhibiting vanishing time-delay ($\protect\phi =0$) with respect
to it. We consider a bonding combination of $1s$ atomic orbitals, and take
the intensity of the attosecond-pulse train to be $I_{\mathrm{h}}=I_{l}/10$.
The remaining parameters are the same as in the previous figures. Panel (a):
spectra computed employing the modified version of the two-center prefactors
in the length form [Eq. ({\protect\ref{modifieddip}})] and single-atom
saddle-point equations, compared to that obtained using modified
saddle-point equations. Panel (b): Contributions from individual scattering
scenarios, i.e., from $|M_{j\protect\nu }|^{2}$, with $j=1,2$ and $\protect%
\nu =1,2$. The interference minima are indicated by arrows in the figure.}
\label{atto1}
\end{figure}

\begin{figure}[tbp]
\begin{center}
\includegraphics[width=8cm]{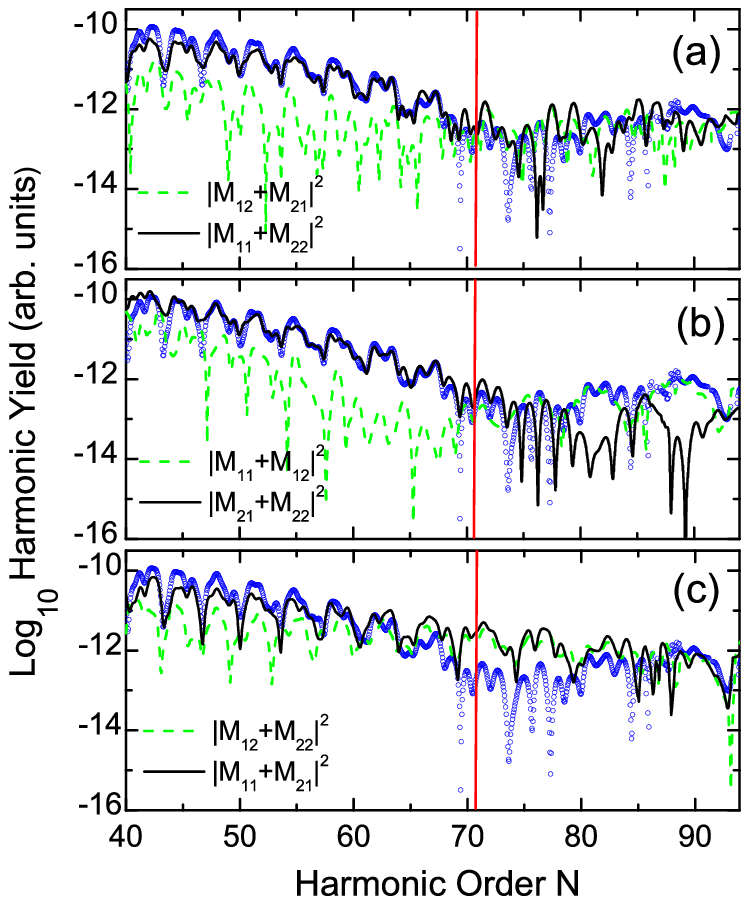}
\end{center}
\caption{(Color Online) Contributions to the high-harmonic yield from the
quantum interference between different types of orbits, for the same field
and molecular parameters as in Fig.~7. Panel (a): Contributions from
topologically similar scattering scenarios, i. e., contributions from $%
|M_{11}+M_{22}|^{2}$ and $|M_{12}+M_{21}|^{2}$. Panel (b): Contributions
from orbits starting at the same center. Panel (c): Contributions from
orbits ending at the same center. For comparison, the overall spectra are
displayed as the blue symbols in the figure. The interference minimum is
indicated by the vertical lines near $\Omega =71\protect\omega $.}
\label{atto2}
\end{figure}

In Fig. \ref{atto1}, we depict the spectra obtained for a diatomic molecule
subjected to such a field, assuming either a two-center prefactor and the
single-center saddle-point equation (\ref{saddelt}), or the modified saddle
point equations (\ref{moddelt}) [Fig. \ref{atto1}.(a)]. Both computations
exhibit a minimum near the harmonic frequency $\Omega =71\omega $, in
agreement with Eq. (\ref{maxmin}). If only the contributions $|M_{j\nu
}|^{2} $ from the individual scattering scenarios are taken, such a minimum
is absent [Fig. \ref{atto1}.(b)]. Therefore, it is due to interference
effects between different sets of orbits. One should note that, in contrast
to the purely monochromatic case, all contributions exhibit the same order
of magnitude. This is due to the fact that, if the attosecond pulses are
present, the electron is being ejected in the continuum with roughly the
same probability, regardless of the center it left from.

The precise role of the various recombination scenarios is illustrated in
Fig. \ref{atto2}. For clarity, we concentrate on the plateau region around
the interference minimum. The main difference observed, with regard to the
purely monochromatic case, is that the overall shape of the spectrum, and
consequently its minimum, is due to the collective interference of several
types of orbits. This is in contrast to the previous results, for which they
were caused by the processes starting at a center $C_{j}$ and ending at
different centers, i.e., $|M_{jv}+M_{jj}|^{2}$, with $\nu \neq j$ and $\nu
,j $ $=(1,2)$. Indeed, a modulation is even present for the probability $%
|M_{11}+M_{22}|^{2}$ involving only one-center scenarios. This is shown in
Fig. \ref{atto2}.(a), and contradicts the previously made assumption that
such features are due to the interference between topologically different
sets of orbits. In fact, it seems that the absence of overall maxima and
minima for the one-center contributions, in the purely monochromatic case
[Fig. \ref{orbits1}.(a)], is due to the different orders of magnitude for $%
M_{11}$ and $M_{22}$ [c.f. Fig. \ref{orbits2}.(a)].

Furthermore, one needs several different processes in order to obtain the
correct position of the minimum. For instance, in Fig. \ref{atto2}.(a), the
overall spectrum closely follows $|M_{11}+M_{22}|^{2}$ in the low-energy
region. In the vicinity of the minimum and for higher energies, however, it
follows neither such contrbutions nor $|M_{12}+M_{21}|^{2},$ from two-center
processes. In Fig. \ref{atto2}.(b), where we display the contributions $%
|M_{jj}+M_{j\nu }|^{2}$, with $\nu \neq j$ and $\nu ,j$ $=(1,2),$ from the
orbits starting from the same center, the spectrum closely follows $%
|M_{21}+M_{22}|^{2}$ before the minimum, and $|M_{11}+M_{12}|^{2}$ after the
minimum.

The remaining panel [Fig. \ref{atto2}.(c)] depicts the contributions from $%
|M_{\nu \nu }+M_{j\nu }|^{2}$ , with $\nu \neq j$ and $\nu ,j$ $=(1,2)$,
which give the orbits \emph{finishing} at the same center. In this case, the
interference minimum is absent. This is a strong evidence that the relevant
condition for the presence of such features is that the orbits taken into
account end at different centers, instead of being topologically different
(which is the case for both Fig. \ref{atto2}.(b) and Fig. \ref{atto2}.(c)).
Therefore, these results agree with the double-slit picture, which has been
put across in \cite{doubleslit}.

The additional attosecond pulses have the advantage of not introducing
changes in the standard length-gauge SFA formulation. They modify, however,
the physics of the problem, since they provide a different mechanism for the
electron to reach the continuum. Clearly, there is also the possibility of
eliminating the spurious potential energy shifts, by considering a different
version of the SFA, such as the field-dressed length-gauge formulation
proposed in \cite{DM2006}, or the velocity-gauge formulation.
\begin{figure}[tbp]
\begin{center}
\includegraphics[width=9cm]{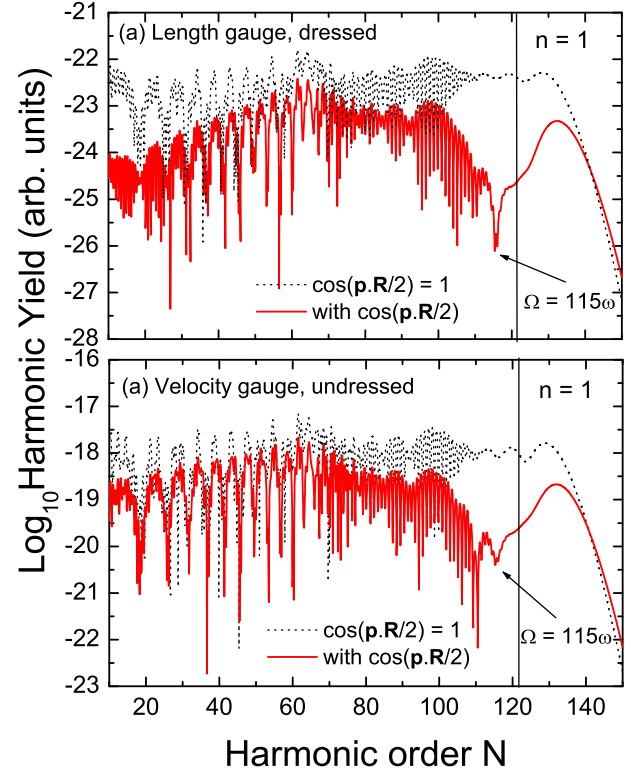}
\end{center}
\caption{(Color Online) High-order harmonic spectra for the same field and
molecular parameters as in Fig. 2, but computed with the field-dressed
modified length-gauge formulation of the strong-field approximation [Panel
(a)], compared to its field-undressed velocity-gauge counterpart [Panel
(b)]. The arrows in the figure indicate the harmonic order for which a
minimum is observed, and the vertical line marks the rough estimate for such
a minimum. The solid and dotted lines correspond to the prefactor (19) and
to the situation for which $\cos \mathbf{p}\cdot \mathbf{R}$/2 has been set
to one, respectively.}
\label{dressed}
\end{figure}

In Fig. \ref{dressed}, we display high-order harmonic spectra for the
field-dressed SFA in the length gauge, and for the field-undressed SFA in
the velocity gauge [panels (a) and (b), respectively]. For simplicity, we
exhibit the results obtained with the modified prefactor (\ref{modifieddip}%
), instead of a modified action. We also provide curves for which only the
cosine term has been set to one, in order to facilitate the identification
of interference effects. As an overall feature, we do not observe the
interference patterns exhibited in the previous figures. Indeed, the curves
with and without the cosine term are very similar. There is, however, a
minimum near $\Omega =115\omega $ for the former case.

For the dressed length-gauge SFA, the above-stated features can be
attributed to the fact that the condition for maxima and minima is now given
by (\ref{maxmin}), with $\mathbf{p}$ instead of $\mathbf{\tilde{p}}=\mathbf{p%
}+\mathbf{A}(t).$ The harmonic frequencies for which they occur can be
easily obtained from condition (\ref{saddle2}), and are given by
\begin{equation}
\Omega =I_{p}+\left[ n^{2}\pi ^{2}/R_{x}^{2}+2n\pi A(t)/R_{x}+A^{2}(t)\right]
/2.  \label{omegadress}
\end{equation}%
An upper bound for $\Omega $ can be estimated as follows. At the electron
return times, the vector potential is roughly $A(t)\lesssim 2\sqrt{U_{p}}.$
This yields, for the parameters in Fig. \ref{dressed}, $\Omega \sim
121\omega $, which is slightly larger than the minimum encountered.

A breakdown of the interference patterns also occurs in the velocity gauge,
for the very same reasons. Indeed, the interference condition for the SFA in
the velocity gauge and for the field-dressed SFA in the length gauge are
identical. This is a direct consequence of the fact that field-free initial
and final electron states in the velocity gauge are gauge-equivalent to the
field-dressed states considered in this paper. This gauge equivalence will
lead to identical recombination form factors $d_{\mathrm{rec}}(\mathbf{p}%
)=\left\langle \mathbf{p}\right\vert \mathcal{O}_{\mathrm{dip}}.\mathbf{e}%
_{x}\left\vert \psi _{0}\right\rangle $. Since the interference conditions (%
\ref{maxmin}) are mainly determined by $d_{\mathrm{rec}}(\mathbf{p})$, they
will be the same. Hence, the harmonic orders for which the maxima and minima
occur are given by Eq. (\ref{omegadress}), and therefore are unrealistically
high. The discrepancies between both yields stem from the form factors $d_{%
\mathrm{ion}}(\mathbf{p})=\left\langle \mathbf{p}\right\vert H_{\mathrm{int}}%
\mathbf{(}t^{\prime }\mathbf{)}\left\vert \psi _{0}\right\rangle $,
which are gauge-dependent.
\begin{figure}[tbp]
\begin{center}
\includegraphics[width=8cm]{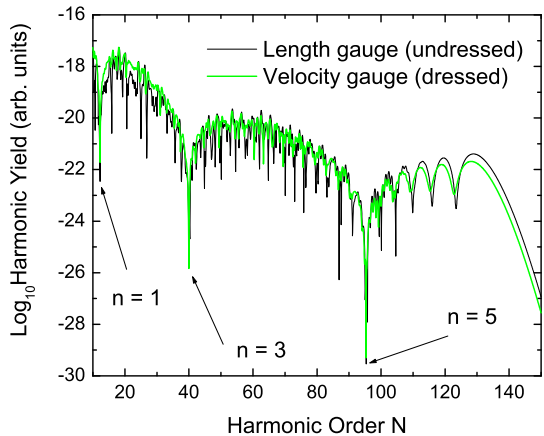}
\end{center}
\caption{(Color Online) High-order harmonic spectra for the same field and
molecular parameters as in Fig. 2, but computed with the field-dressed
modified velocity-gauge formulation of the strong-field approximation. In
order to facilitate the comparison, the undressed length-gauge version of
the SFA is provided, and the velocity-gauge yield has been normalized in
approximately 2 orders of magnitude, to match the length-gauge results. In
both cases, for simplicity, the modified prefactor (18) has been employed,
instead of modified saddle-point equations.}
\end{figure}

One may, however, consider field-dressed states in the velocity gauge, which
are gauge-equivalent to field-free states in the length gauge. This is
achieved by applying the transformation $\chi _{v\leftarrow l}=\exp [i%
\mathbf{A}(t)\cdot \mathbf{r}]$ in the initial and final length-gauge
electronic bound states.\ This leads to a shift $\mathbf{p\rightarrow p+A}(t)
$ on a momentum eigenstate $\left\vert \mathbf{p}\right\rangle $, which is
exactly the opposite shift induced in the field-dressed length-gauge SFA.
This shift has the main consequence that the interference condition (\ref%
{maxmin}) now holds for $\mathbf{\tilde{p}}=\mathbf{p}+\mathbf{A}(t)$, even
in the velocity gauge. The results obtained employing such dressed states in
the velocity-gauge SFA, depicted in Fig. 9, are indeed very similar to those
obtained using its field undressed, length gauge formulation. In fact, we
have observed mainly quantitative differences, due to different prefactors $%
d_{\mathrm{ion}}(\mathbf{\tilde{p}})$ \cite{footndion2}.

One should note, however, that the transformation $\chi _{v\leftarrow l}$
introduces the same additional potential energy shifts in the modified
action as in the undressed length-gauge case. Thus, the price one pays for
recovering the correct interference conditions is the loss of a direct
connection to simple classical models.

\section{Conclusions}

\label{concl}

The present results support the viewpoint that the maxima and minima in the
high-order harmonic spectra of diatomic molecules are due to the
interference of electron orbits finishing at different centers in the
molecule. This seems to hold regardless of whether the electron has been
released in one center at the molecule and recombines with a different
center, or whether it is ejected and returns to the same center. Such
conclusions have been reached by employing modified saddle-point equations,
within the strong-field approximation. These modifications lead to orbits
involving different centers, and are a slightly more refined approach than
the standard procedure of considering single-center saddle-point equations
and modified prefactors.

In this framework, we considered that the electron has been ejected by
tunneling ionization and by an additional attosecond-pulse train, and
compared the similarities and differences from both physical situations. In
the former case, depending from which center the electron is leaving, it
must overcome unequal potential barriers to reach the continuum, whereas in
the latter case it is ejected with roughly equal probability. Especially in
the purely monochromatic case, we observed that the contributions from
orbits starting at one of the centers, namely $C_2$, were much larger than
those from $C_1$, due to a narrower potential barrier. In particular, there
exist potential-energy shifts which are proportional to the electric field
at the ionization time and the internuclear distance, which cause such
differences. It is however noteworthy that the electron excursion times have
been confined to the first half-cycle of the laser field. If the other
half-cycle had been taken, the potential barrier would reverse and the
contributions starting from $C_1$ would be more prominent.

The above-mentioned potential-energy shifts, however, do not possess a
clear-cut physical interpretation. In fact, they are only present in the
standard length-gauge formulation of the strong-field approximation, i.e.,
if field-free bound states are taken when the electron is ejected and
recombines, and are the source of several problems. For instance, they
reflect the fact that the SFA is translation-dependent. Moreover, due to
their existence, it is difficult to establish an immediate connection
between this approach and the classical equations of motion of an electron
in the laser field.

On the other hand, it seems that such shifts are necessary in order to
obtain the correct energy position of the maxima and minima in the HHG
spectra. In fact, an improved formulation of the SFA, in which the influence
of the laser field is included in the electron bound states, restores its
translation invariance, provides an unproblematic classical limit \cite%
{DM2006}, but yields incorrect energy positions for the interference
patterns. This discrepancy is related to the fact that the field dressing
alters the double-slit interference conditions (\ref{maxmin}). A similar
absence of interference features has been reported very recently in Ref.
\cite{SSY2007}, for HHG computations using a field-dressed version of the
SFA in the length gauge.

Finally, when employing different gauges, from our results it is
clear that the dressing of the initial and the final states plays a
far more important role than the different interaction Hamiltonians
$H_{\mathrm{int}}(t^{\prime })$. If the dressing is applied
consistently so that the the electronic bound states are
gauge-equivalent, the interference patterns will remain the same.
This is
due to the fact that the interference condition (\ref%
{maxmin}) will then remain invariant. For instance, the spectra
obtained in the velocity-gauge SFA with undressed states are very
similar to those computed in the length gauge with field dressed
states. The same holds for the undressed length-gauge, and the
dressed velocity-gauge SFA spectra. In the two former cases, there
is a breakdown of the interference patterns, as compared to the
field-undressed length gauge SFA. However, such patterns can be
restored in the velocity gauge, by dressing the electronic bound
states appropriately.

\acknowledgements We would like to thank L. E. Chipperfield, R. Torres, J.
P. Marangos, and H. Schomerus for useful discussions, and W. Becker for
calling Ref. \cite{DM2006} to our attention. We are also grateful to the
Imperial College and to the University of Stellenbosch for their kind
hospitality. This work has been financed by the UK EPSRC (Advanced
Fellowship, Grant no. EP/D07309X/1).


\begin{thebibliography}{99}
\bibitem{footngauge} In the length or velocity gauge, this interaction is
given by $H_{I}^{L}(t^{\prime })=\mathbf{r}.\mathbf{E}(t^{\prime })$ or $%
H_{I}^{V}(t^{\prime })=[\mathbf{p}+\mathbf{A}(t^{\prime })]^{2}/2$,
respectively.
\bibitem{spreading} Since the bound state with which the returning electron
recombines is highly localized, a broader wave packet means a less
pronounced overlap upon recombination and, consequently, weaker harmonics.
\end{thebibliography}
\end{document}